\documentclass[twocolumn,aps,prb,showpacs,preprintnumbers,amsmath,amssymb,superscriptaddress,10pt]{revtex4-1}

\usepackage{graphicx}
\usepackage{dcolumn}
\usepackage{bm}
\usepackage{dcolumn}
\usepackage{braket}
\usepackage{amsmath}
\usepackage{array}
\begin{document}


\title{The Kerr rotation in the unconventional superconductor Sr$_2$RuO$_4$}


\author{Martin Gradhand}
\email{m.gradhand@bristol.ac.uk}
\affiliation{H . H. Wills Physics Laboratory, University of Bristol, Tyndall Ave, BS8-1TL, UK}

\author{Karol I. Wysokinski}
\affiliation{Institute of Physics, M. Curie-Sk\l odowska University, Radziszewskiego 10, PL-20-031 Lublin, Poland}

\author{James F. Annett}
\affiliation{H . H. Wills Physics Laboratory, University of Bristol, Tyndall Ave, BS8-1TL, UK}

\author{Balazs L. Gy\"{o}rffy}
\affiliation{H . H. Wills Physics Laboratory, University of Bristol, Tyndall Ave, BS8-1TL, UK}


\date{\today}

\begin{abstract}
The interpretation of Kerr rotation measurements in the superconducting phase of Sr$_2$RuO$_4$ is a controversial topic. Both intrinsic and extrinsic mechanisms have been proposed, and it has been argued that the intrinsic response vanishes by symmetry. We focus on the intrinsic contribution and clarify several conflicting results in the literature. On the basis of symmetry considerations and detailed calculations we show that the intrinsic Kerr signal is not forbidden in a general multi-band system but has a rich structure in the near infrared regime. We distinguish different optical transitions determined by the superconducting gap (far infrared) and the inter orbital coupling of the normal state (near infrared). We argue that the low frequency transitions do not contribute to the Hall conductivity while only the inter-orbital transitions in the near infrared regime contribute. Finally, we discuss the difficulties to connect the calculations for the optical Hall conductivity to the experimental measurement of the Kerr angle. We will compare different approximations which might lead to conflicting results.         
\end{abstract}


\maketitle

\section{Introduction}\label{sec:intro}
The measurements of the finite Kerr rotation angle in the superconducting phase of Sr$_2$RuO$_4$ \cite{Xia2006, Kapitulnik2009} attracted a large interest since they are believed to hold the decisive proof for time reversal symmetry (TRS) breaking in this unconventional superconductor. The experimental observation generated an enormous body of theoretical work trying to explain the findings qualitatively and even quantitatively. In the beginning the work was focused on intrinsic mechanisms \cite{Yip1992,Read2000,Yakovenko2007,Mineev2007,Lutchyn2008,Roy2008} neglecting effects arising from impurity scattering. These studies referred to various mechanisms such as collective modes and finite size of the laser spot or where proven to be incorrect later on. A common feature between all these approaches was the restriction to single band models. Subsequently, it was argued by several authors that the Kerr effect has to be zero in the intrinsic and homogeneous superconductors which is the crucial point we want to address here. Those arguments are correct for single band models but we will contend that for multiorbital systems the general arguments do not hold. Only recently, two groups \cite{Taylor2012, Wysokinski2012} showed that in multi band models an intrinsic effect can exist indeed and is of the order of magnitude as the experimentally found value \cite{Xia2006}. Almost immediately these results were criticized by Mineev.~\cite{Mineev2012} He argued, on the basis of symmetry arguments, that the intrinsic Kerr effect has to vanish even in a multi orbital approach. We will address this criticism in detail. Finally, despite the fact that both groups~\cite{Taylor2012, Wysokinski2012} performed calculations based on very similar models their findings are surprisingly different. The considered major contributions are in a significantly different frequency range. We will highlight these discrepancies and show what caused them. Here we analyze both approaches and compare the results in detail. 

A quite different route to explain the experiments is based on impurity scattering.~\cite{Kim2008,Goryo2008,Lutchyn2009} These approaches turned out to be as successful in describing the experiment as the intrinsic mechanism. However, whether the experiments are caused by an intrinsic or an extrinsic mechanism can hardly be decided on the basis of the existing data. It will be crucial to predict features relying on one or the other mechanism which can be proven experimentally. In the following we will concentrate on the intrinsic mechanism. 

This article will be organized in six major sections. After this introduction, we start with the description of our numerical approach to calculate the Kerr rotation angle based on the evaluation of the optical conductivity tensor. In the following section we will introduce the three orbital and three dimensional model for the superconducting state of Sr$_2$RuO$_4$ in detail.\cite{Annett2002} The numerical results will be presented in Sec.~IV. Finally, we compare our results to the approach of Ref.~\onlinecite{Taylor2012} and show its similarities and differences. This discussion will include arguments why the criticism of Mineev \cite{Mineev2012} does not apply to the considered models. In the last section we are going to make contact to the experimental observation of the Kerr angle and highlight the uncertainties in reliable predictions. 

\section{Theoretical method}\label{sec:theory}

In our approach \cite{Wysokinski2012} we follow the analysis of Capelle, Gross, and Gy\"{o}rffy \cite{Capelle1997,Capelle1998} for the magneto optical dichroism in superconductors. We will only discuss their main results relevant for our numerical implementation and the following discussion. The frequency dependent Kerr angle is given by \cite{Lutchyn2009} 
\begin{equation}\label{Eq:Kerr}
\theta_K=\frac{1}{\epsilon_0\omega}\text{ Im}\frac{\sigma_{xy}(\omega)}{n(\omega)(n^2(\omega)-1)}\ \text{,}
\end{equation}
where $n(\omega)$ is the complex refraction coefficient and $\sigma_{xy}(\omega)$ is the complex optical Hall conductivity. In general, the existence or the absence of the dichroic signal is determined by $\sigma_{xy}(\omega)$ and so we focus on that quantity in the following. However, for quantitative predictions of the Kerr angle in comparison to experiments the knowledge of the complex refraction coefficient is needed as well. We will comment on different approximations later. According to Refs.~\onlinecite{Capelle1997,Capelle1998} the real and imaginary part of the optical Hall conductivity can be expressed as
\begin{widetext}
\begin{equation}\label{Eq:Im_sigmaxy}
\text{Im}\left[\sigma_{xy}(\omega)\right]=\frac{\pi e^2}{2\omega Vm^2}\sum\limits_{n,n^\prime, {\bf k}}f(E_{n{\bf k}})\left[1-f(E_{n^\prime{\bf k}})\right]\left(\left|\braket{\Psi_{n^\prime {\bf k}} |H^{\dagger}_I(\epsilon_L)|\Psi_{n{\bf k}}}\right|^2-\left|\braket{\Psi_{n^\prime {\bf k}}|H^{\dagger}_I(\epsilon_R)|\Psi_{n{\bf k}}}\right|^2\right)\delta(E_{n^\prime{\bf k}}-E_{n{\bf k}}-\hbar\omega)
\end{equation}
\begin{equation}\label{Eq:Re_sigmaxy}
\text{Re}\left[\sigma_{xy}(\omega)\right]=\frac{e^2 \hbar}{Vm^2}\sum\limits_{n,n^\prime, {\bf k}}f(E_{n{\bf k}})\left[1-f(E_{n^\prime{\bf k}})\right]\frac{\left(\left|\braket{ \Psi_{n^\prime {\bf k}} |H^{\dagger}_I(\epsilon_L)|\Psi_{n{\bf k}}}\right|^2-\left|\braket{\Psi_{n^\prime {\bf k}}|H^{\dagger}_I(\epsilon_R)|\Psi{n{\bf k}}}\right|^2\right)}{(E_{n^\prime{\bf k}}-E_{n{\bf k}})^2-\hbar^2\omega^2}
\end{equation}
\end{widetext}
where the interaction Hamiltonian for the absorption of the electromagnetic wave $H^\dagger_I(\epsilon_{L/R})$ is given by
\begin{equation}\label{Eq:Hint} 
H^\dagger_I(\boldsymbol{\epsilon}_{L/R})=
\frac{1}{i}\left(
\begin{array}{cc}
\boldsymbol{\epsilon}^\ast_{L/R}\cdot \hat{\bf p}&0\\
0&-\boldsymbol{\epsilon}^{\ast}_{L/R}\cdot \hat{\bf p}^{\ast}
\end{array}\right)\ .
\end{equation}
Here $\hat{\bf p}$ is the momentum operator $\hbar/i \boldsymbol{\nabla}_{\bf r}$ and $\boldsymbol{\epsilon}_{L/R}=(1,\pm i,0)/\sqrt{2}$. If we consider $-\hat{\bf p}^\ast=\hat{\bf p}$ this lead us to 
\begin{equation}\label{Eq:Hint_2} 
H^\dagger_I(\boldsymbol{\epsilon}_{L/R})=
\frac{1}{i}\left(
\begin{array}{cc}
\boldsymbol{\epsilon}^\ast_{L/R}\cdot \hat{\bf p}&0\\
0&\boldsymbol{\epsilon}^{\ast}_{L/R}\cdot \hat{\bf p}
\end{array}\right)\ .
\end{equation}
The wavefunctions
$\braket{{\bf r}|\Psi_{n{\bf k}}}=e^{i{\bf kr}}\left(
u_{n{\bf k}}({\bf r}),v_{n{\bf k}}({\bf r})
\right)^T $
are Bloch like solutions of the Bogoliubov-de Gennes (BdG) equations
\begin{equation}\label{Eq:BdG}
\left(
\begin{array}{cc}
\hat{H}_{\bf k}({\bf r})& \hat{\Delta}({\bf r})\\
\hat{\Delta}^{\dagger}({\bf r})&-\hat{H}_{-{\bf k}}^{\ast}({\bf r})
\end{array}
\right)\left(
\begin{array}{c}
u_{n{\bf k}}({\bf r})\\
v_{n{\bf k}}({\bf r})
\end{array}
\right)
=E_{n{\bf k}}
\left(
\begin{array}{c}
u_{n{\bf k}}({\bf r})\\
v_{n{\bf k}}({\bf r})
\end{array}
\right)
\end{equation}
with the ${\bf k}$-dependent lattice periodic normal state Hamiltonian $\hat{H}_{\bf k}=e^{-i{\bf kr}}\hat{H}({\bf r)} e^{i{\bf kr}}$. If we apply the identity $\hat{\boldsymbol p}/m=\hat{\boldsymbol v}=e^{i{\bf kr}}\nabla_{\bf k}H_{\bf k}/\hbar e^{-i{\bf kr}}$ in Eq.~(\ref{Eq:Hint}) we get
\begin{equation}\label{Eq:Hint2}
H^\dagger_I(\boldsymbol{\epsilon}_{L/R})=\frac{1}{i}e^{i{\bf kr}}\frac{m}{\sqrt{2}\hbar}\left(\hat{H}^x_{\bf k}\mp i\hat{H}^y_{\bf k}\right)e^{-i{\bf kr}}
\end{equation}
with
\begin{equation}\label{Eq:Hint3}
\hat{H}^{x,y}_{\bf k}=
\left(
\begin{array}{cc}
\partial_{k_{x,y}}\hat{H}_{\bf k}({\bf r})&0\\
0&\partial_{k_{x,y}}\hat{H}_{\bf k}({\bf r})
\end{array}\right)\ \text{.}\\
\end{equation}
Combining Eqs.~(\ref{Eq:Im_sigmaxy}) and (\ref{Eq:Hint2}) we can express the optical conductivity in terms of the periodic part of the Bloch function $\braket{{\bf r}|{n{\bf k}}}=\left(
u_{n{\bf k}}({\bf r}),v_{n{\bf k}}({\bf r})
\right)^T $
\begin{widetext}
\begin{eqnarray}
\text{Im}\left[\sigma_{xy}(\omega)\right]&=&\frac{\pi e^2}{2\omega V\hbar^2}\sum\limits_{n,n^\prime, {\bf k}}f(E_{n{\bf k}})\left[1-f(E_{n^\prime{\bf k}})\right]\delta(E_{n^\prime{\bf k}}-E_{n{\bf k}}-\hbar\omega)\times\nonumber\\
&&\text{Im}\left[\braket{ {n^\prime {\bf k}} |H^x_{\bf k}|{n {\bf k}}}\braket{ {n {\bf k}}|H^y_{\bf k}|{n^\prime {\bf k}}}-\braket{ {n^\prime {\bf k}} |H^y_{\bf k}|{n {\bf k}}}\braket{ {n {\bf k}} |H^x_{\bf k}|{n^\prime {\bf k}}}\right]\ \text{.}
\end{eqnarray}
\end{widetext}
This looks like the expression used in Ref.~\onlinecite{Wysokinski2012} but is crucially different. The first point is that $\hat{H}_{\bf k}({\bf r})$ is the ${\bf k}$-dependent Hamiltonian but not the tight-binding Hamiltonian. More crucially $H^{x,y}_{\bf k} $ is a diagonal matrix with two equivalent entries in contrast to Ref.~\onlinecite{Wysokinski2012} where the lower entry had opposite sign to the upper term. This will cause large differences in the numerical results, namely no low frequency contributions. In the following we will rewrite the expression in terms of the tight-binding Hamiltonian under consideration.

A fairly general expansion of the Bloch wavefunction in a superconductor takes the form
\begin{equation}\label{Eq:TB_exp}
\braket{{\bf r}|\Psi_{n{\bf k}}}=
\sum_{L,{\bf R}}e^{i{\bf kR}}\left(
\begin{array}{c}
u_{n}({\bf k})\\
v_{n}({\bf k})
\end{array}
\right)_{L}\Phi_{L}({\bf r}-{\bf R})
\end{equation}
where $L$ represents the local orbital quantum number within the unit cell while ${\bf R}$ refers to the site within the periodic lattice.  Exploiting this expansion we can rewrite the BdG equation~(\ref{Eq:BdG}) as, 
\begin{equation}\label{Eq:BdG_TB}
\left(
\begin{array}{cc}
H({\bf k})& \hat{\Delta}({\bf k})\\
\hat{\Delta}^{\dagger}({\bf k})&-H^{\ast}(-{\bf k})
\end{array}
\right)\left(
\begin{array}{c}
u_{n}({\bf k})\\
v_{n}({\bf k})
\end{array}
\right)
=E_{n{\bf k}}
\left(
\begin{array}{c}
u_{n}({\bf k})\\
v_{n}({\bf k})
\end{array}
\right)
\end{equation}
where all entries in the operator on the left hand side are matrices in the local orbital space which were included in the expansion~(\ref{Eq:TB_exp}). Accordingly, the eigensolutions are vectors in the orbital space. The matrix elements can be expressed as
\begin{eqnarray}
H_{L,L^\prime}({\bf k})&=&\sum_{\bf R}e^{i {\bf k} {\bf R}}\int\limits_{UC}d^3r\Phi^\ast_L({\bf r})H({\bf r})\Phi_{L^\prime}({\bf r}-{\bf R})\ \text{,}\\
\Delta_{L,L^\prime}({\bf k})&=&\sum_{\bf R}e^{i {\bf k} {\bf R}}\int\limits_{UC}d^3r\Phi^\ast_L({\bf r})\Delta({\bf r})\Phi_{L^\prime}({\bf r}-{\bf R})\ \text{.}
\end{eqnarray}
The integration runs over one unit cell (UC) only. These are standard results using the on-site approximation, i.e. no overlapping of neighboring basis functions $\int d^3r\Phi^\ast_L({\bf r}-{\bf R}^\prime)\Phi_{L^\prime}({\bf r}-{\bf R}) =\delta_{{\bf RR}^\prime}\delta_{LL^\prime}$. Now we need to express the interaction Hamiltonian of Eq.~(\ref{Eq:Hint_2}) in terms of the tight binding basis,  which gives the result
\begin{widetext}
\begin{equation}\label{Eq:Hint_TB}
\braket{n^\prime{\bf k} |H^{\dagger}_I(\epsilon_{L/R})|n{\bf k}}=\frac{\boldsymbol{\epsilon}_{L/R}^\ast}{i}\frac{m}{\hbar}\left(
\begin{array}{c}
u_{n^\prime}({\bf k})\\
v_{n^\prime}({\bf k})
\end{array}
\right)^\dagger
\left(
\begin{array}{cc}
\nabla_{\bf k}H({\bf k})& 0\\
0&\nabla_{\bf k}H({\bf k})
\end{array}
\right)\left(
\begin{array}{c}
u_{n}({\bf k})\\
v_{n}({\bf k})
\end{array}
\right)\ \text{.}
\end{equation}
\end{widetext}
To derive that formula we used $\hat{\bf p}=-i\frac{m}{\hbar}\left[\hat{\bf r}\hat{H}-\hat{H}\hat{\bf r}\right]$, the completeness relation $\sum\limits_{L,{\bf R}}\Phi_L({\bf r}-{\bf R})\Phi^\ast_L({\bf r}^\prime-{\bf R})=\delta({\bf r}-{\bf r}^\prime)$, and most importantly the assumption of a vanishing dipole moment within the unit cell
\begin{equation}
\int d^3{\bf r}\ \Phi^\ast_L({\bf r}-{\bf R})\ \hat{\bf r}\ \Phi_{L^\prime}({\bf r}-{\bf R}^\prime)={\bf R}\delta_{LL^\prime}\delta_{{\bf R}{\bf R}^\prime}\ \text{.}
\end{equation}
To summarize this derivation it needs to be said that Eq.~(\ref{Eq:Hint_TB}) is very similar to the one derived in Ref.~\onlinecite{Wysokinski2012}. However, it is crucially different since it omits the minus sign in front of the lower ${\bf k}$ derivative of the ${\bf k}$-dependent tight-binding Hamiltonian. In the following we will show that for the low frequency region of the optical conductivity this implies dramatic changes. 
 
\section{Tight binding model for $\text{Sr}_2\text{RuO}_4$} \label{sec:TB_model}
The results for the optical Kerr effect in the superconducting phase of   Sr$_2$RuO$_4$ which we are going to present in section~\ref{sec:numerics} are based on a three-orbital and three-dimensional tight binding model. This model was introduced in the literature\cite{Annett2002,Annett2003,Wysokinski2003} already and relies on the following main facts. First, it describes accurately the experimentally found three sheet Fermi surface \cite{Mackenzie1996,Bergemann2000} and the cyclotron masses.\cite{Mackenzie1998} Second, it is a minimal model in the interaction parameters which reproduces the experimentally found specific heat \cite{Nishizaki2000,Annett2002} quantitatively. Finally, and following from the previous point, it includes horizontal line nodes of the gap function and reproduces the superfluid density.\cite{Annett2002,Annett2003} Here, we will only state the final structure of the model, with slight adjustments of the interaction parameters to account for the improved numerical accuracy. In the following we will apply this model, without any further changes, to calculate the optical Kerr effect, finally comparing it to the experimental observations. 
As mentioned already, our model is based on three Ru-4d orbitals ($d_{xy}$, $d_{xz}$, $d_{yz}$) which in the following we will denote as ($a$, $b$, $c$), respectively. For the proper description of the normal state electronic bandstructure we consider:
\begin{eqnarray*}
H_{aa}({\bf k}) &= &\varepsilon_a+2t\left(\textrm{cos}\,k_x+ \textrm{cos}\,k_y\right)+4t^\prime\,\textrm{cos}\,k_x\,\textrm{cos}\,k_y\\
H_{bb}({\bf k}) &= &\varepsilon_b+2\left(t_b^x\,\textrm{cos}\,k_x + t_b^y\,\textrm{cos}\,k_y\right)\\&+&8t^\perp_{b}\,\textrm{cos}\,\frac{k_x}{2}\, \textrm{cos}\,\frac{k_y}{2}\,\textrm{cos}\,\frac{k_z}{2}c\\
H_{cc}({\bf k}) &= &\varepsilon_c+2\left(t_c^x\,\textrm{cos}\,k_x + t_c^y\,\textrm{cos}\,k_y\right)\\ &+ & 8t^\perp_{c}\,\textrm{cos}\,\frac{k_x}{2}\,\textrm{cos}\,\frac{k_y}{2}\, \textrm{cos}\,\frac{k_z}{2}c\\
H_{ab}({\bf k}) &= &8t^\perp_{ab}\,\textrm{cos}\,\frac{k_x}{2}\, \textrm{sin}\,\frac{k_y}{2}\,\textrm{sin}\,\frac{k_z}{2}c\\
H_{ac}({\bf k}) &= &8t^\perp_{ac}\,\textrm{sin}\,\frac{k_x}{2}\,\textrm{cos}\,\frac{k_y}{2}\, \textrm{sin}\,\frac{k_z}{2}c\\
H_{bc}({\bf k}) &= &4t_{bc}\,\textrm{sin}\,k_x\,\textrm{sin}\, k_y+8t^\perp_{bc}\,\textrm{ sin}\,\frac{k_x}{2}\,\textrm{sin}\,\frac{k_y}{2}\, 
\textrm{cos}\,\frac{k_z}{2}c
\end{eqnarray*}
For symmetry reasons we can impose the following conditions on the introduced parameters
\begin{equation*}
\varepsilon_b=\varepsilon_c,\
t_c^y=t_b^x,\ 
t_c^x=t_b^y,\
t^\perp_{c}=t^\perp_{b}, \text{ and }
t^\perp_{ab}=t^\perp_{ac}\ ,   
\end{equation*}
which reduces the number of parameters to 10. Using this set of parameters we are able to fit the experimentally found cyclotron masses \cite{Mackenzie1998}, Fermi surface areas \cite{Mackenzie1996} and the most pronounced corrugations found in z direction.\cite{Bergemann2000} 
The in plane lattice constant is set to one and the out of plane lattice constant is c=3.279. For the superconducting state the minimal model takes the explicit form
\begin{eqnarray}\label{Eq:Gapfunction}
\Delta^{\uparrow\downarrow}_{aa}({\bf k}) &= &\eta^{x}_{aa}\,\textrm{sin}\,k_x+ \eta^{y}_{aa}\,\textrm{sin}k_y\nonumber\\
\Delta^{\uparrow\downarrow}_{bb}({\bf k}) &= &\eta^{x}_{bb}\, \textrm{sin}\,\frac{k_x}{2}\,\textrm{cos}\frac{k_y}{2}\,\textrm{cos}\, \frac{k_z}{2}c \nonumber\\ 
&+& \eta^{y}_{bb}\,\textrm{cos}\,\frac{k_x}{2}\,\textrm{sin}\, \frac{k_y}{2}\,\textrm{cos}\,\frac{k_z}{2}c\nonumber \\
\Delta^{\uparrow\downarrow}_{cc}({\bf k}) &= &\eta^{x}_{cc}\, \textrm{sin}\,\frac{k_x}{2}\,\textrm{cos}\,\frac{k_y}{2}\, \textrm{cos}\,\frac{k_z}{2}c\\
&+& \eta^{y}_{cc}\, \textrm{cos}\,\frac{k_x}{2}\,\textrm{sin}\, \frac{k_y}{2}\,\textrm{cos}\,\frac{k_z}{2}c\nonumber \\
\Delta^{\uparrow\downarrow}_{bc}({\bf k}) &= &\eta^{x}_{bc}\, \textrm{sin}\,\frac{k_x}{2}\,\textrm{cos}\,\frac{k_y}{2}\,\textrm{cos}\, \frac{k_z}{2}c\nonumber \\
&+& \eta^{y}_{bc}\, \textrm{cos}\,\frac{k_x}{2}\,\textrm{sin}\, \frac{k_y}{2}\, \textrm{cos}\,\frac{k_z}{2}c\nonumber
\end{eqnarray}
where the order parameters are given by
\begin{widetext}
\begin{eqnarray*}
\eta^{x}_{aa} &=\ U\ &\sum_n\int \text{d}^3\textbf{k}\left\{u^{\uparrow}_{a,n}({\bf k})(v^{\downarrow}_{a,n}({\bf k}))^\ast +u^{\downarrow}_{a,n}({\bf k})(v^{\uparrow}_{a,n}({\bf k}))^\ast\right\}\textrm{sin}\,k_x\ [1-2f(T,E_n)]\\
\eta^{x}_{bb}&= 4U^\prime&\sum_n\int \text{d}^3\textbf{k}\left\{u^{\uparrow}_{b,n}({\bf k})(v^{\downarrow}_{b,n}({\bf k}))^\ast +u^{\downarrow}_{b,n}({\bf k})(v^{\uparrow}_{b,n}({\bf k}))^\ast\right\} \textrm{sin}\,\frac{k_y}{2}\,\textrm{cos}\,\frac{k_y}{2}\, \textrm{cos}\,\frac{k_z}{2}c\ [1-2f(T,E_n)]\\
\eta^{x}_{cc}&= 4U^\prime&\sum_n\int \text{d}^3\textbf{k}\left\{u^{\uparrow}_{c,n}({\bf k})(v^{\downarrow}_{c,n}({\bf k}))^\ast +u^{\downarrow}_{c,n}({\bf k})(v^{\uparrow}_{c,n}({\bf k}))^\ast\right\} \textrm{sin}\,\frac{k_y}{2}\,\textrm{cos}\,\frac{k_y}{2}\, \textrm{cos}\,\frac{k_z}{2}c\ [1-2f(T,E_n)]\\
\eta^{x}_{bc}&= 4U^\prime&\sum_n\int \text{d}^3\textbf{k}\left\{u^{\uparrow}_{b,n}({\bf k})(v^{\downarrow}_{c,n}({\bf k}))^\ast +u^{\downarrow}_{c,n}({\bf k})(v^{\uparrow}_{b,n}({\bf k}))^\ast\right\} \textrm{sin}\,\frac{k_y}{2}\,\textrm{cos}\,\frac{k_y}{2}\, \textrm{cos}\,\frac{k_z}{2}c\ [1-2f(T,E_n)].\\
\end{eqnarray*}
\end{widetext}
In combination with a set of symmetry induced relations for the order parameters
\begin{equation*}
\eta_{aa}^y=i\,\eta_{aa}^x,\ \eta_{bb}^y=i\,\eta_{cc}^x,\ \eta_{cc}^y=i\,\eta_{bb}^x,\ \eta_{bc}^y=-i\,\eta_{bc}^x,\
\end{equation*}
and the gap function
\begin{equation*}
\Delta^{\uparrow\downarrow}_{ji}({\bf k})=\Delta^{\downarrow\uparrow}_{ij}({\bf k}) 
\end{equation*}
this fully determines the gap function entering the BdG equation (\ref{Eq:BdG_TB}).
The actual parameters used for the following calculations are summarized in Table~\ref{Tab:para} and are essentially equivalent to those introduced in the literature.\cite{Annett2003} It should be mentioned that due to an improved numerical accuracy it was necessary to slightly adjust the interaction parameters $U$ and $U^\prime$ to fix the transition temperature at 1.5K.
\begin{table*}[t]
\begin{ruledtabular}
\begin{tabular}{c|c|c|c|c|c|c|c|c|c|c|c|c}
$\varepsilon_a$ & $\varepsilon_b$ & $t$ & $t^\prime$ & $t^x_b$ & $t^y_b$ &  $t^\perp_b$ & $t^\perp_{ab}$ & $t^\perp_{ac}$ & $t_{bc}$ & $t^\perp_{bc}$  & $U $ & $U^\prime$\\
&&&&&&&&&&&&\\
\hline
&&&&&&&&&&&&\\
$-131.8$ & $-132.22$ & $-81.62$ & $-36.73$ & $-109.37$ & $-6.56$ & $0.262$ & $-1.05$ & $-1.05$ & $-8.75$ & $-1.05$ & $37.63$ & $50.69$ \\
&&&&&&&&&&&&\\
\end{tabular}
\end{ruledtabular}
\caption{The numerical parameters for the tight-binding model of Sr$_2$RuO$_4$ and the interaction parameters describing the superconducting phase in units of $\text{meV}$.} 
\label{Tab:para}
\end{table*}  
For numerical calculations, solving the BdG equations self-consistently and for the optical conductivity, we used a very dense ${\bf k}$-point mesh with $480\times 480\times 48$ points along the Brillouin zone basis vectors (1,0,1/c), (0,1,1/c), and (0,0,2/c).

\section{Numerical Results}\label{sec:numerics}
\newlength{\LL} \LL 0.8\linewidth
\begin{figure}[ht]
\includegraphics[width=\LL]{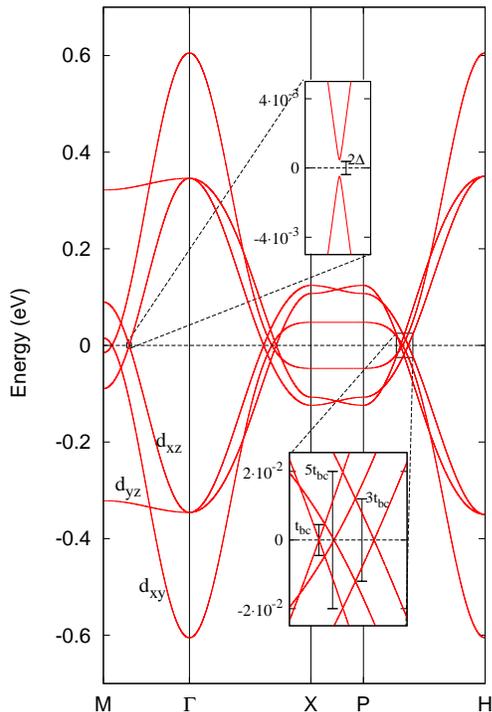}
\caption{Quasi particle bandstructure of the considered three band model for the superconducting phase of Sr$_2$RuO$_4$ along high symmetry lines. As insets we highlight different possible transitions between electron and hole bands. The order of magnitude for the energies of transitions between the same orbital character is set by the gap approximately as $2\Delta_{bb} = 7.6*10^{-4}\ \text{eV}$ which is in the far infrared frequency regime. Transitions between different orbital characters are given by  $t_{bc} \approx 0.009\ \text{eV}$ $(xy\rightarrow yz)$, $5t_{bc} = 0.044\ \text{eV}$ $(xz\rightarrow yz)$, and $3t_{bc} = 0.026\ \text{eV}$ $(xz\rightarrow yz)$ respectively\label{fig:band}}
\end{figure}
After we have set up the theoretical background in Sec.~\ref{sec:theory} and defined the actual model in Sec.~\ref{sec:TB_model} we will now discuss the numerical results for the optical conductivity. To set a framework for the following discussion Fig.~\ref{fig:band} shows the quasiparticle bandstructure at $T=0\ \text{K}$ along high symmetry lines. 
We have three bands dominantly related to the orbital character $d_{xy}$, $d_{xz}$, and $d_{yz}$. The contributions to the optical conductivity which will be discussed in the following can be viewed as transition between the positive and negative solutions of the quasi-particle spectrum separated by the opening of the superconducting gap. The two energy scales for possible transitions conserving the crystalline momentum ${\bf k}$ are set by the superconducting gap of the order of $2\Delta_{bb}({\bf k}=(2\pi,0,0))=7.6*10^{-4}\ \text{eV}$ and the inter orbital hopping between $xz$ and $yz$ orbitals $t_{bc} = 0.9*10^{-2}\ \text{eV}$. Since the first one is related to transitions between electron and hole bands of the same predominant orbital character we will refer to them as intra-orbital transition while we label the latter one as inter-orbital transitions. The inter-orbital transitions can be further separated into three distinct transitions according to $xy\rightarrow xz$, $xy\rightarrow yz$, and $xz\rightarrow yz$ of which $xy\rightarrow yz$ are lowest in energy.
The energies and frequencies of the intra orbital and inter orbital transitions are separated by two orders of magnitude, connected to the far infrared and near infrared spectrum, respectively.
\newlength{\LLL} \LLL 0.7\linewidth
\begin{figure}[ht]
\includegraphics[width=\LLL, angle=-90]{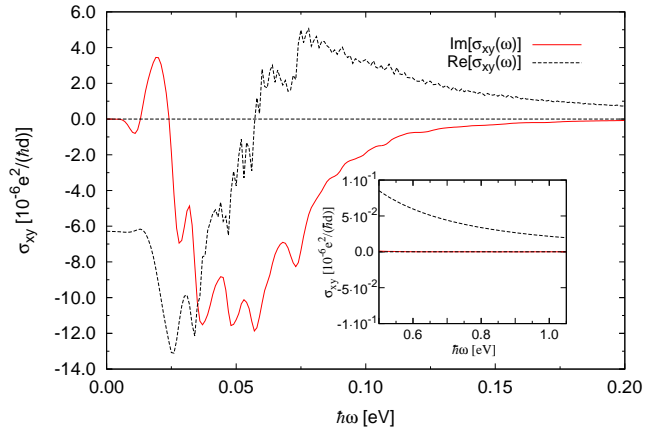}%
\caption{The real and imaginary part of the optical conductivity according to Eq.~(\ref{Eq:Im_sigmaxy}). The inset shows the high frequency results in the range of the experiments ($\omega_{exp}=0.8\ \text{eV}$).\cite{Xia2006} The conductivities are given in units of $e^2/ \hbar /d$, where $d=6.37\mathring{A}$ is the c-axes interlayer spacing.\label{fig:Kerr}}
\end{figure}
In Fig.~\ref{fig:Kerr} we present our results for the real and imaginary part of the optical conductivity according to Eqs.~(\ref{Eq:Re_sigmaxy}) and (\ref{Eq:Im_sigmaxy}). 
The Figure shows the near infrared frequency region since the signal is zero below the onset around $0.008\ \text{eV}$. It corresponds to the energy scale given by the inter-orbital transitions as introduced above. Both real and imaginary part show a rich structure with pronounced features between $0.01$ and  $0.1\ \text{eV}$. The vanishing of the far infrared (low energy) signal is distinct from the findings in a previous publication of some of the authors.~\cite{Wysokinski2012} This is related to the different signs discussed in relation to the interaction Hamiltonian of Eqs.~(8) and (9) in this work. The inset of Fig.~\ref{fig:Kerr} shows the frequency range around 0.8 eV where the experimental effect was observed.\cite{Xia2006} The imaginary part almost vanishes in that regime while the real part remains finite and becomes $\text{Re}[\sigma_{xy}(\omega=0.8\ \text{eV})=3.4*10^{-8}\text{e}^2/(\hbar d)$ 

\begin{figure}[ht]
\includegraphics[width=\LLL, angle=-90]{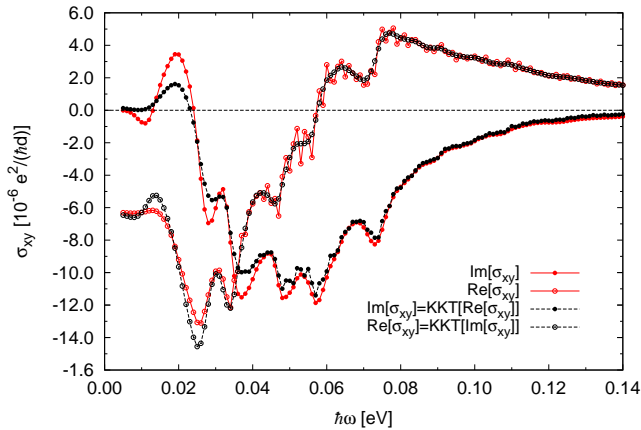}%
\caption{The real and imaginary part of the optical conductivity according to Eqs.~(\ref{Eq:Im_sigmaxy}) and (\ref{Eq:Re_sigmaxy}) in comparison to the same quantities as obtained from the Kramers-Kronig transformation. The very good agreement validates the numerical accuracy of the calculations. \label{fig:Kerr_KKT}}
\end{figure}

For a numerical validation of the applied method we performed Kramers-Kronig transformations for both the imaginary and the real part. We compare these in Fig.~\ref{fig:Kerr_KKT} to the direct calculations. Clearly, the agreement is very good owing to the fact that the numerical precision and stability of the calculations are satisfactory. 

In the following section we will compare our approach and the numerical results to other approaches which show conflicting results at the first
glance. We will point out the reasons for the discrepancies and resolve the contradictions.  
\section{Discussion 1: Comparison to existing approaches}\label{sec:discuss1}
Almost at the same time when some of the authors published their original work for the Kerr effect in Sr$_2$RuO$_4$\cite{Wysokinski2012} another group presented a very similar approach leading to quite different results. Taylor and Kallin~\cite{Taylor2012} found a finite signal using a two-dimensional two-band model which, according to their results showed a finite signal only in the near infrared region. This work was criticized by Mineev~\cite{Mineev2012} who argued about the vanishing intrinsic Kerr signal in Sr$_2$RuO$_4$ even for a two band model.

We will start the comparison of our results with Taylor and Kallin.~\cite{Taylor2012} The corresponding model was based on the $d_{xz}$($b$) and $d_{yz}$($c$) orbitals only and was limited to two dimensions, i.e. no interlayer coupling and $k_z=0$. In addition, compared to Eq.~(\ref{Eq:Gapfunction}), they neglected the inter-orbital gap $\Delta^{\uparrow\downarrow}_{bc}({\bf k})$, set $t_b^y=t_c^x=0$ and $\eta_{bb}^y=\eta_{cc}^x=0$. This simplified model does not fully describe the normal state Fermi surface, the cyclotron masses, nor the experimentally observed specific heat accurately. The model has no line nodes in the gap function and therefore will have an exponential specific heat for low temperature in contradiction to the quadratic dependence found in the experiment.\cite{Nishizaki2000,Annett2002} Despite of these differences it is illustrative to analyze this model in more detail. 
\begin{figure}[ht]
\includegraphics[width=\LL]{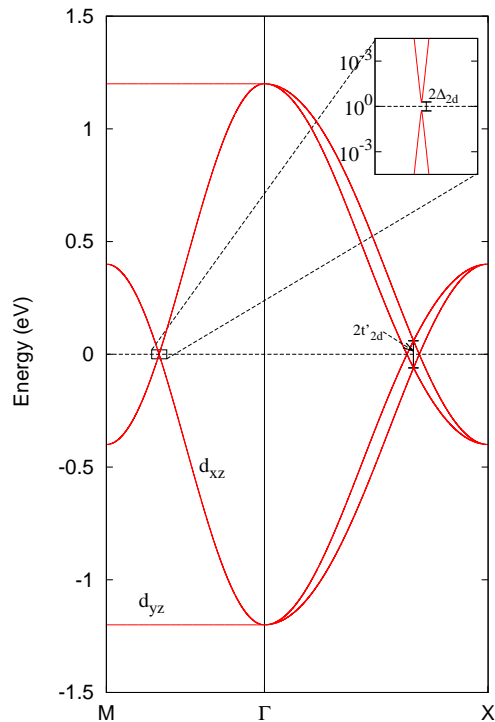}%
\caption{Quasi particle bandstructure of the reduced two-dimensional and two-band model for the superconducting phase of Sr$_2$RuO$_4$ as applied by Taylor and Kallin.\cite{Taylor2012} Possible transition between electron and hole bands of the same orbital character separated by $2\Delta = 4.0*10^{-4}\ \text{eV}$ and the different orbital character ($xz$, $yz$) separated by 0.12 eV are marked in the inset and the main figure respectively      \label{fig:band_2d}}
\end{figure}
In Figure~\ref{fig:band_2d} we show the corresponding two-dimensional quasiparticle bandstructure using exactly the same model parameters as introduced in Ref.~\onlinecite{Taylor2012}. Comparing this to our three-dimensional model as shown in Fig.~\ref{fig:band} the overall band width is much larger. This is caused by parameters fitted to density functional theory (DFT) calculations in contrast to our fit to experimental results. It is well known that typical DFT calculations overestimate the bandwidth drastically.~\cite{Mackenzie1996} The orders of magnitude for the different transitions are set by $2\Delta_{2d}^x=3.0\ast 10^{-4}\ \text{eV}$ and $2t^\prime=0.12\ \text{eV}$, very similar to what we found for the three-dimensional model.

\begin{figure}[ht]
\includegraphics[width=\LLL, angle=-90]{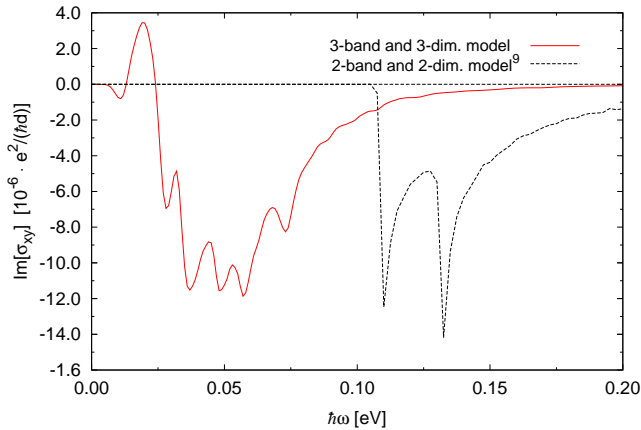}%
\caption{The imaginary part of the optical conductivity according to Eq.~(\ref{Eq:Im_sigmaxy}) comparing the three dimensional and three orbital model of the current work to the two dimensional and two orbital model as used by Taylor and Kallin.~\cite{Taylor2012} The agreement is reasonable. The three band model has a much richer structure due to the existence of many more possible transitions.~\cite{Annett2003} 
\label{fig:Kerr_Kallin_model}}
\end{figure}

\begin{figure}[ht]
\includegraphics[width=\LLL, angle=-90]{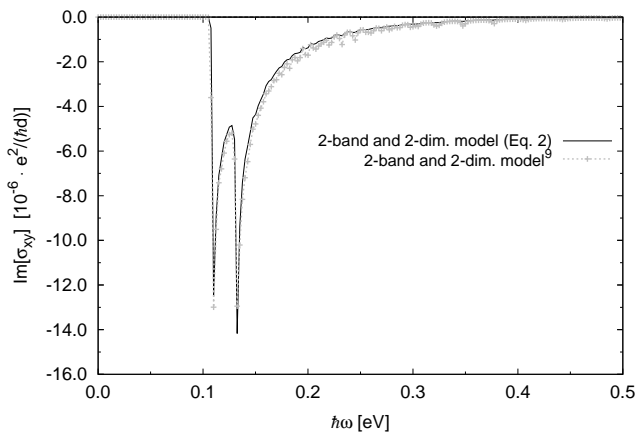}
\caption{The imaginary part of the optical conductivity for the two band and two dimensional tight-binding model of superconducting Sr$_2$RuO$_4$ according to Taylor and Kallin.~\cite{Taylor2012} The black solid line shows the result applying  Eq.~(\ref{Eq:Im_sigmaxy}) of the current work while the grey broken line uses Eq.~(11) of Ref.~\onlinecite{Taylor2012}.    
\label{fig:Kerr_Kallin}}
\end{figure}

In the following we use the Taylor and Kallin model\cite{Taylor2012} to calculate the optical conductivity according to Eq.~(\ref{Eq:Im_sigmaxy}) and compare the results to our full three-dimensional model. 
This comparison is shown in Fig.~\ref{fig:Kerr_Kallin_model} for the near infrared, high frequency, region. The results are very similar with only a shift to lower frequencies. This is caused by the significantly larger bandwidth of the two-dimensional model as pointed out above. 
Here, we would like to highlight two further important points. First, the pronounced positive feature at around $\hbar\omega=0.015\ \text{eV}$ in the three-band model is determined by transitions between $xy$ and $yz$ orbitals which do not exist in the reduced model. Secondly, the original discrepancy between both approaches \cite{Wysokinski2012,Taylor2012} concerning the existence of low frequency contributions of the order of the gap function is resolved in this work. There is no optical signal in the far infrared region.  
To show the equivalence of our numerical solution of Eq.~(\ref{Eq:Im_sigmaxy}) and the derivation of Ref.~\onlinecite{Taylor2012} we show this comparison in Fig.~\ref{fig:Kerr_Kallin}. The grey broken line shows the result according to Eq.~(11) of Ref.~\onlinecite{Taylor2012} while the black solid curve is the solution of Eq.~(\ref{Eq:Im_sigmaxy}) in this work but using the same tight-binding model as Taylor and Kallin.\cite{Taylor2012}
The agreement is perfect.

\begin{figure}[ht]
\includegraphics[width=\LLL, angle=-90]{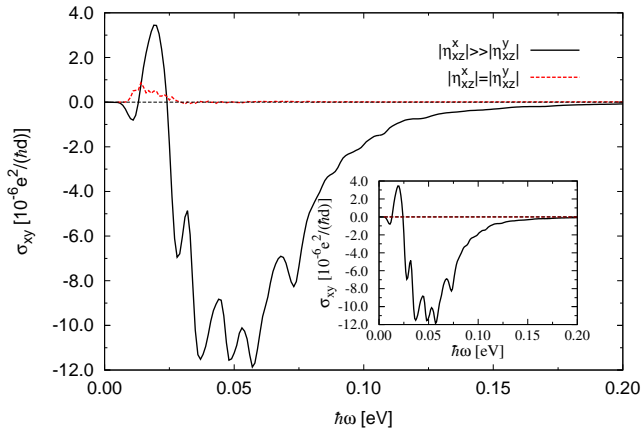}%
\caption{The imaginary part of the optical conductivity according to Eq.~(\ref{Eq:Im_sigmaxy}) using our general model in comparison to a calculation artificially obeying condition~(\ref{Eq:Mineev_cond}). The inset shows the same comparison including one further condition, namely $t_{ab}^\perp=t_{ac}^\perp=0$, which decouples the $xy$ orbitals from the $xz$ and $yz$ orbitals. In that case the imaginary part vanishes in accordance to Ref.~\onlinecite{Mineev2012}.   
\label{fig:Kerr_Mineev}}
\end{figure}

With that we turn our attention to even more crucial results in literature, arguing that the intrinsic optical conductivity vanishes in translational invariant superconductors.\cite{Mineev2012} The arguments were based on fairly general symmetry considerations for the structure of the gap function. In his work Mineev~\cite{Mineev2012} focused on the model as used in Ref.~\onlinecite{Taylor2012} and derived two crucial statements: 1. A two dimensional representation of the gap function is essential to correctly describe the symmetry of the system. 2. Using the adequate two dimensional representation leads inevitably to a vanishing optical conductivity. However, in his second argument his considerations were not accurate. The expressions for the gap functions which he was using had the form 
\begin{eqnarray*}
\Delta^{\uparrow\downarrow}_{bb}({\bf k})&=&(\eta^x_{bb}\, \textrm{sin}\, k_x+\eta^y_{bb}\, \textrm{sin}\, k_y)\\
&\text{and}&\\
\Delta^{\uparrow\downarrow}_{cc}({\bf k})&=&(\eta^x_{cc}\, \textrm{sin}\, k_x+\eta^y_{cc}\, \textrm{sin}\, k_y)
\end{eqnarray*}
where the order parameter amplitudes $\eta^x_{bb}$, $\eta^y_{bb}$, $\eta^x_{cc}$, and $\eta^y_{cc}$ were determined by the free energy expansion. The crucial condition he was using are the equalities
\begin{eqnarray}\label{Eq:Mineev_cond}
\left|\eta^x_{bb}\right| &=&\left|\eta^y_{bb}\right|\nonumber\\
&\text{and}&\\
\left|\eta^x_{cc}\right| &=&\left|\eta^y_{cc}\right|\ .\nonumber
\end{eqnarray}
However, this is not valid in general since interactions of $d_{xz}$ orbitals in $x$ and $y$ direction are not related by any symmetry arguments at all. In fact, solving the BdG equations self-consistently we find the corresponding terms to differ by one order of magnitude. This disproves his argument since it was based on a subtle cancellation of different contributions which fails if the equalities above do not hold. To underline this finding once more, we performed an artificial calculation where we fixed the model to obey the above conditions (\ref{Eq:Mineev_cond}). The result is shown in Fig.~\ref{fig:Kerr_Mineev} in comparison to our self-consistent model. Evidently, the optical conductivity almost vanishes imposing conditions (\ref{Eq:Mineev_cond}). Only at the onset of the signal, created via the transitions between $xy$ $yz$ orbitals a small contribution remains. However, the $d_{xy}$ orbitals were neglected in Ref.~\onlinecite{Mineev2012} so could not contribute at all. Here, the signal is created via the normal state hopping between $xy$, $xz$ and $xy$, $yz$ orbitals with distinct gap parameters. In a two dimensional (single plane) model there is no $d_{xy}$ to $d_{yz}$ ($d_{xz}$) hopping by symmetry, but in three dimensions this is possible. For that reason the subtle cancellation does not take place and we find a finite optical conductivity. To confirm this argument we provide as inset of Fig.~\ref{fig:Kerr_Mineev} the result of a calculation where we omit, in addition to condition~(\ref{Eq:Mineev_cond}), the normal state coupling between $xy$ and $xz$ ($yz$) orbitals, $t_{ab}^\perp=t_{ac}^\perp=0$. Including this further assumption the optical conductivity vanishes identically which is reassuring and agrees with the result of Mineev.~\cite{Mineev2012}
Concluding this section we would like to stress ones more that the arguments for a vanishing intrinsic Kerr effect~\cite{Read2000, Mineev2012} are strictly valid for a single band model. However, in a more general multi-band system a finite Kerr can be observed if time-reversal symmetry or inversion symmetry are broken. 
\section{Discussion 2: Comparison to experimental result}\label{sec:discus21}
The final part of this paper is dedicated to the comparison with experimental results.\cite{Xia2006} This is not as straight forward as it seems due to the fact that the measured quantity is the Kerr angle in contrast to the optical conductivity we discussed so far. To make contact between both we need to evaluate Eq.~(\ref{Eq:Kerr}) with the so far unknown complex refraction coefficient $n(\omega)=\sqrt{\epsilon(\omega)}$. The permeability can be expressed as
\begin{equation}
\epsilon(\omega)=\epsilon_\infty+\frac{i}{\omega}\frac{\sigma_{xx}(\omega)}{\epsilon_0}
 \end{equation} 
which leaves us with the unknown complex longitudinal optical conductivity $\sigma_{xx}(\omega)$ and $\epsilon_{\infty}$. In the literature various approaches were proposed and tested to derive these numbers from the existing experiments.\cite{Katsufuji1996, Xia2006, Lutchyn2009, Wysokinski2012, Taylor2012} However, these attempts were relying on more or less accurate approximation of the experimentally found longitudinal optical conductivity by a Drude model. In general, this would read
\begin{equation}\label{Eq:sxx}
\sigma_{xx}(\omega)=-\frac{\epsilon_0\omega^2_{pl}}{i (\omega+i\gamma)}
\end{equation}
where the scattering rate $\gamma$ would be in principle frequency dependent. Both, $\gamma$ and the plasma frequency $\omega_{pl}$ have to be approximated. Here, we would like to show that these approximations are good enough to roughly estimate the magnitude of the possible Kerr angle but are by no means reliable to predict an accurate quantitative result. One of the problems is the fact that the frequency of the experiment deriving the Kerr angle is actually close to the plasma frequency of the system. Despite that it is questionable how reliably the optical response of this multi band metal/superconductor can be described by a general Drude model at all. In Fig~\ref{fig:Kerr_angle} we compare different approaches to address that point.      
\newlength{\LLLL} \LLLL 0.9\linewidth
\begin{figure}[ht]
\includegraphics[width=\LLLL]{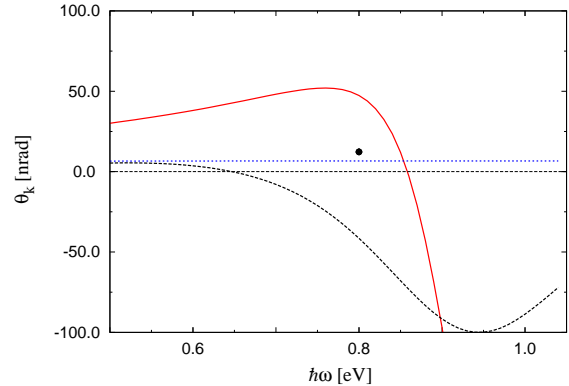}%
\caption{The estimated Kerr angle using the optical Hall conductivity as shown in the inset of Fig.~\ref{fig:Kerr} and different approximations for the permeability $\epsilon(\omega)$. The solid red curve is based on Eq.~(\ref{Eq:sxx}) using the parameters $\epsilon_\infty=10$, 
$\hbar\omega_{pl}=2.9\ \text{eV}$, and $\gamma=0.4$ derived in Refs.~\onlinecite{Taylor2012,Lutchyn2009,Katsufuji1996}. The broken black line is exploiting the same model but using $\gamma=0.2$. The dotted blue line uses the assumption of $\omega_{pl}/\sqrt{\epsilon_\infty}>>\omega $ and $\hbar\omega_{pl}=4.5\ \text{eV}$ as discussed in Ref.~\onlinecite{Wysokinski2012}. Finally, the black circle is derived from the experimental data presented in Ref.~\onlinecite{Katsufuji1996} leading to parameters for the permeability as $\text{Re}[\epsilon(\hbar\omega=0.8\ \text{eV})]=-6.09$ and $\text{Im}[\epsilon(\hbar\omega=0.8\ \text{eV})]=7.43$ as discussed in the text. The resulting Kerr angle is $12\ \text{nrad}$. The experimental Kerr angle at $\hbar\omega=0.8\ \text{eV}$ is in the range of $60$ to $90\ \text{nrad}$.   
\label{fig:Kerr_angle}}
\end{figure}
The three lines in Fig.~\ref{fig:Kerr_angle} refer to existing approaches based on Eq.~\ref{Eq:sxx} exploiting different sets of parameter derived from an experimental work.\cite{Katsufuji1996} However, all these approaches are based on the assumption that it is reasonable to describe the longitudinal optical response by means of a Drude model. Even so these approaches are very similar the figure shows that slight variations in the parameters can drastically change the estimated Kerr angle. To avoid this difficulty we derived the real and imaginary part of the refraction index $n(\hbar\omega=0.8\ \text{eV})$ from the experiment\cite{Katsufuji1996} explicitly. Here, we used the connection between the reflectivity and the refraction index $R=|(n-1)/(n+1)|^2$. We took the real part of the optical conductivity from Fig.~1 of Ref.~\onlinecite{Katsufuji1996} to be $\text{Re}[\sigma_{xx}(\hbar\omega=0.8\ \text{eV})\approx 800\ \Omega^{-1}\text{cm}^{-1}$ and the reflectivity $R(\hbar\omega=0.8\ \text{eV})\approx 0.6$ from these numbers the real and imaginary part of n can be derived to be $\text{Im}[n(\hbar\omega=0.8\ \text{eV})\approx 7.43$ and $\text{Re}[n(\hbar\omega=0.8\ \text{eV})\approx -6.09$. The resulting Kerr angle is $\Theta_K(\hbar\omega=0.8\ \text{eV})=12\ \text{nrad}$. It is similar in magnitude to the other approximations but reduces the uncertainty about the values of the necessary parameters. Nevertheless, all these approaches remain rough approximations and thorough experimental and theoretical work on the complex valued longitudinal conductivity are highly desirable.  
\section{Summary} 
In conclusion we have presented a thorough analysis of the mechanism and the theoretical and numerical description of the optical Kerr effect in superconducting Sr$_2$RuO$_4$. We clarified the principle of the existence or not of the intrinsic effect and point to the crucial ingredients for any model describing the problem. Furthermore, we resolve various contradictions between earlier works especially about the existence of low frequency contributions to the spectrum and the magnitude of the effect. Due to the fact that our full three band description takes care of the experimentally observed line nodes in the superconducting state, accurately describes the measured Fermi surface properties, and is able to account for many quantities such as specific heat and superfluid density quantitatively we are reassured that the quantitative estimation of the Kerr angle is reasonable. However, the comparison to the Taylor and Kallin model suggests that the presence or absence of gap nodes is not essential to the Kerr signal. Finally, we argue that further theoretical as well as experimental investigations are crucial to make further quantitative contact between theory and experiment. Of special importance is here the proper description of the longitudinal component of the optical conductivity which was treated so far by approximations relying on the applicability of the generalized Drude model. Ultimately, measurements of the Kerr angle for a larger frequency range are important to be able to make final conclusion about the intrinsic or extrinsic nature of the observed optical Kerr effect.
\begin{acknowledgments}
This work was carried out using the computational facilities of the Advanced Computing Research Centre, University of Bristol - http://www.bris.ac.uk/acrc/.
M.G acknowledges financial support from the DFG via a research fellowship (GR3838/1-1). We thank Daniel Fritsch for helpful discussions concerning the numerical implementation of the tight-binding BdG equation.
\end{acknowledgments}

\bibliography{../../My_Collection}

\end{document}